\documentstyle[12pt,aasms4]{article}
\begin{document}

\def\cen{\centerline}
\def\nid{\noindent}
\def\sqdeg{$\sqcup\mskip-12.0mu \sqcap^\circ$ }
\def\ie{{\it i.e. }}
\def\etal{{ et al. }}
\def\simgt{\hbox{\rlap{\raise 0.425ex\hbox{$>$}}\lower 0.65ex\hbox{$\sim$}}}
\def\simlt{\hbox{\rlap{\raise 0.425ex\hbox{$<$}}\lower 0.65ex\hbox{$\sim$}}}

\title{The Effect of Weak Gravitational Lensing on 
the Angular Distribution of Gamma-Ray Bursts}
\author{L. L. R. Williams}
\affil{Institute of Astronomy, Madingley Road, Cambridge CB3 0HA, UK}
\authoremail{llrw@ast.cam.ac.uk}

\begin{abstract}
If Gamma-Ray Bursts (GRBs) are cosmologically distributed standard 
candles and are associated with the luminous galaxies, then the 
observed angular distribution of all GRBs is altered due to weak
gravitational lensing of bursts by density inhomogeneities. The amplitude 
of the effect is generally small. For example, if the current catalogs
extend to $z_{max}\sim 1$ and we live in a flat $\Omega=1$ Universe,
the angular auto-correlation function of GRBs will be enhanced by
$\sim 8\%$ due to lensing, on all angular scales.
For an extreme case of $z_{max}= 1.5$ and 
($\Omega$, $\Lambda$)=(0.2, 0.8), an enhancement of $\sim 33\%$ is
predicted. If the observed distribution of GRBs is used in the future
to derive power spectra of mass density fluctuations on large angular
scales, the effect of weak lensing should probably be taken into account.
\end{abstract}


\section{Introduction}

After more than two decades of GRB observations their physical nature
and distances remain largely a mystery. The distribution of bursts 
is isotropic in the sky (Quashnock 1996, Tegmark \etal 1996). 
This observation rules out Galactic disk/bulge models, but leaves 
Galactic halo (Lamb 1995) and cosmological models as possible scenarios.
The Spectroscopy Detectors on board the Compton Gamma Ray Observatory
detected no convincing spectral features to help determine GRB 
redshifts (Palmer \etal 1993). Searching for counterparts of GRBs at 
other
wavelengths (optical, IR, X-rays) could help determine their distances,
however none were found within the error boxes of GRBs (Shaefer 1993,
but see Irwin \& \.Zytkow 1994).
In fact, the lack of any bright optical or IR galaxies in the error 
boxes of 10 brightest GRBs indicates that the bursts must originate at
distances greater than 1 Gpc (Shaefer 1990).

If the bursts are extragalactic, they may be expected to trace luminous 
matter in the Universe. For example, if they are associated with 
coalescing black holes or neutron stars, their rate would be roughly
proportional to the density of normal stellar populations. 
In particular, the bursts should have 
clustering properties similar to those of the galaxies.
All attempts to find angular correlation of GRBs with themselves
(Blumenthal \etal 1993, Quashnock 1996, Tegmark \etal 1996), 
and with nearby cosmological structures 
(Hartmann \etal 1996, Nemiroff \etal 1993) 
have so far failed. These non-detections can be used to place lower 
limits on the ``effective" distance to the GRB population.  Using
GRBs from the 1B Catalog, Blumenthal \etal (1993) derive a lower limit 
of $z\sim 0.05$.
If typical faint GRBs were any closer one would detect a positive
angular correlation signal, since a smaller effective
distance means less smearing of the clustering signal due to projection
effects. Based on the data from the 3B Catalog,
Quashnock (1996) derives a minimum sampling distance of $z=0.25$, at
95\% confidence level. Based on the lack of any Supergalactic 
anisotropies, Hartmann \etal (1996) derive a minimum sampling distance 
of $z\sim 0.065$.

All auto- and cross- correlation analyses with the BATSE Catalog are 
restricted to angles greater than $\sim 5^\circ$.
Smaller separations cannot be probed reliably because of the low surface
density of the bursts (roughly 1 GRB per 40 square degrees in the 3B
Catalog), and more importantly, because of the large positional
errors in the burst locations. 
Positions of individual GRBs in the BATSE Catalog suffer from random and
systematic errors. The random errors range from 0.5 to 15 degrees 
depending on the flux of the GRB, while the systematic errors are 
about 4 degrees (Graziani \& Lamb 1996).

In this paper we investigate the effect of weak gravitational lensing 
on the distribution of GRBs. If the GRBs are standard candles and trace 
the luminous matter, should we expect to see weak lensing effects in the
angular distribution of bursts of all fluxes?
 
To show how weak lensing could affect the observed distribution of GRBs
one can explore the following toy model. Imagine a positive density 
inhomogeneity  (a clump of galaxies) located somewhere between us and
the typical faint GRBs. The clump itself will produce an increase in the
number of GRBs seen in that direction. It will also act as a 
lens for the background matter. The effects of lensing can be calculated
in a straightforward fashion. Due to flux 
magnification, the maximum redshift of observed GRBs will be extended, 
and the surface number density of background GRBs will be diluted; 
therefore, that direction in the sky
will have an excess of bursts due to the galaxy clump and lensing 
flux magnification, as well as a deficit due to lensing area distortion. 
How do the strengths of these effects compare? In Section 2 we will see
that lensing can produce changes in the distribution of GRBs at the
level of several percent. Due to projection effects, however, the signal 
will be smeared out and is not expected to be seen in the current data 
sets. 

Before proceeding, we would like to note that the influence of strong 
lensing on GRBs has been considered in the literature, see
Paczy\'nski (1986, 1987). In general, if GRBs extend to $z\sim 2$, 
the fraction of strongly lensed bursts should be comparable to that of
QSOs, \ie $\sim4$ in every 500 GRBs should be multiply imaged 
(see Maoz \& Rix 1993). Therefore there
could be several multiply imaged GRBs in the current BATSE Catalog.

\section{The Model}

\subsection{Assumptions about the GRB population}

Throughout this paper we assume that GRBs are cosmological in origin, 
all the bursts have identical spectral properties, and identical
luminosities. We assume that the luminosity and number density of GRBs
do not evolve with redshift. There is no convincing physical reason for 
such simplifications, 
except that a model based on these very simple assumptions correctly
predicts the log$N$-log$f$ distribution of sources over 3 decades in
flux (Mao \& Paczy\'nski 1992).

In particular, bright bursts have a cumulative number counts distribution
with slope of $-{3\over 2}$, expected from a uniform spatial 
distribution of nonevolving sources in flat space. In addition,
the well documented roll-off in the observed number counts of GRBs
at faint fluxes is competely consistent with the geometry of an expanding
Universe. The minimum distance to the faintest GRBs, derived using
best fits to the log$N$-log$f$ curves, is $z\simgt 1$ (Wickramasinghe
\etal 1993, Mao \& Paczy\'nski 1992). These distances
are in agreement with the lower limits derived using 
auto- and cross- correlation analyses (see Introduction). 
One must remember, however, that the assumptions of no-evolution and
single intrinsic burst luminosity are not necessary, and may not be 
sufficient to describe the current GRB data.

We assume, following Mao and Paczy\'nski (1992), that the spectral slope
of the GRBs is $a=1$, such that $L_{\nu}d\nu \propto \nu^{a-1} d\nu$, 
and the flux from a burst at $z$ is given by 
$$f\propto (1+z)^a/[4\pi D_L(z)^2],\eqno(1)$$ 
where $D_L$ is the luminosity
distance.  Spectral slope is important for K-corrections and 
determination of observed fluxes, since the frequency range of the
BATSE detector corresponds to different energy bands depending on the 
source redshift. Since the bursts are
assumed to be standard candles, the maximum redshift of observed GRBs,
$z_{max}$, is directly related to the faintest detectable flux,
$f_{min}$, through equation (1).

Cosmological time dilation affects GRBs in two ways: it changes the
durations of individual bursts, as well as the rate of events seen by 
the observer. It is interesting to point out that the time dilation of 
event durations may have already been detected (Norris \etal 1994, 
Norris \etal 1996,  but see Brainerd 1994). In the present work time
dilation of individual events will not be important. Only the reduction 
of GRB rate with redshift is relevant for us here.

We assume that the rate of production of GRBs is proportional to 
the density of luminous matter. Also, the mass distribution on
large scales is traced by the light distribution. This assertion is
consistent with the observations of the large scale flows; the mass 
density, reconstructed from the peculiar velocity field of the galaxies, 
agrees reasonably well with the galaxy distribution (Dekel 1994).

We consider three sets of cosmological parameters, 
($\Omega$, $\Lambda$)=(1, 0), (0.2, 0), and (0.2, 0.8), and a
Hubble constant of 75 km~s$^{-1}$~Mpc$^{-1}$ where appropriate.

\subsection{Effects of weak lensing}

It is instructive to study lensing effects due to a single isolated 
density inhomogeneity. Imagine looking at a patch of the sky
of angular diameter $\theta$. GRBs in that patch will originate from
a range of redshifts, or from a conical `tube' with its apex at the
observer and extending up to $z_{max}$. Let the mass distribution within
the tube be completely smooth. The rate of GRBs (of all fluxes) from 
that tube, as seen by an observer, is proportional to $F_u$, where
subscript $u$ stands for `unlensed'. Let us now
place a mass concentration (a `clump') at a redshift $z_l$ within 
this tube, centered on the tube's axis. Let the clump contribute an
additional rate of GRBs proportional to $F_c$. The clump will weakly
lense the background sources, thereby altering the rate of GRBs
originating in the tube. The altered rate of GRBs from the tube
alone is proportional to $F_l$, where subscript $l$ 
stands for `lensed'.

The average, background subtracted, surface mass density of the clump,
$\Sigma_c(r_c)$, is assumed to be proportional to $r_c^{-\beta}$, 
within radius $r_c$. This relation can be normalized such that 
$\Sigma_c r_c^\beta=C$, a constant. A reasonable
value for $\beta$ is probably around 1, corresponding to a singular
isothermal sphere, or 0.8, which is the slope of the observed angular 
correlation function of galaxies. The exact values of $\beta$ and $C$ 
are not important, as will be shown later (see equation [7]).

Since the angular scales that can be probed with the GRB population are 
of the order of a few degrees or more, the cosmological structures  
defined by such scales are correspondingly large in extent and 
represent small deviations from the average mass density of the 
Universe.  To illustrate this, consider an angular scale of 10 degrees, 
\ie about the
smallest angular scale that can be reliably sampled with the BATSE data,
given both systematic and random positional errors. At a nearby distance
of $z\sim 0.025$, $10^\circ$ corresponds to $\sim 13h^{-1}$Mpc. On this
scale, the rms scatter in $\delta \rho / \rho$ is less than one.
Therefore, all the  mass clumps considered here have low surface mass 
densities, and are all sub-critical in the lensing sense, 
\ie $\Sigma_c \ll \Sigma_{crit}$. Here, 
$\Sigma_{crit}={{c^2}\over{4\pi G}}{{D_{os}}\over{D_{ol}D_{ls}}}$ is
the critical surface mass density for lensing, and $D$'s are the angular
diameter distances between the source, lens, and observer. 
Strong lensing is unimportant on such scales. Moreover, weak lensing is 
quite weak indeed; therefore it is sufficient to approximate magnification 
$A$ of a lens with $\kappa=\Sigma_c/\Sigma_{crit}$, by 
$A^{-1}\approx(1-\kappa)^2\approx(1-2\kappa)$.

We will now evaluate the expected rates of GRBs, $F_c$, $F_u$, and 
$F_l$. Assuming the production of GRBs traces luminous matter, the clump 
will generate GRBs at a rate proportional to $M_c/(M/L)_c$, where $M_c$ 
and $(M/L)_c$ are its
mass and mass to light ratio. The rate of bursts is then given by,
$$F_c(z_l)\propto {1\over{(M/L)_c}}\Sigma_c r_c^2 (1+z_l)^2 (1+z_l)^{-1}=
{1\over{(M/L)_c}}C r_c^{2-\beta} (1+z_l) .\eqno(2)$$
The time dilation term, $(1+z_l)^{-1}$, is due to the reduced rate
of GRBs originating at cosmological redshifts. The physical surface 
density of a density inhomogeneity increases as $(1+z_l)^2$.
Later, we will need to express $r_c$ in terms of its angular size, and
distance, $r_c\approx \theta~D(z_l)$. The small angle approximation,
assumed here, is valid for angles up to a few tens of degrees. The
angular correlation function of GRBs on scales larger than that is 
probably not relevant. Thus, small angle approximation used here,
and also in equations (3)-(5) below, is valid for our purposes.

Similarly, in the tube, the rate of bursts is also
proportional to the density of luminous matter, 
${(M/L)_U}^{-1} \int \rho_U(z) dV_{tube}(z)$. 
Here, $\rho_U(z)$ and $(M/L)_U$ are the average
density and mass to light ratio of the Universe, and $V_{tube}$ is the
volume of the tube. The rate of GRBs from the tube is given by,
$$F_u(z_{max})\propto{{\rho}\over{(M/L)_U}}\int_0^{z_{max}}
[\theta D(z)]^2{{cdt}\over{dz}}(1+z)^3 (1+z)^{-1} dz,\eqno(3)$$
where $D(z)$ is the angular diameter distance, and $\rho=\rho_U$($z$=0) 
is the present day average mass density of the Universe. 

Weak lensing due to a density inhomogeneity will have two effects on 
the observed sources: (1) it will extend the redshift of
the visible GRBs beyond $z_{max}$, to $z_{max}^\prime$; and (2) 
it will dilute the surface density of the bursts located beyond $z_l$.
Thus, including lensing effects, the rate of GRBs from the tube is 
given by,
$$F_l(z_l,z_{max}^\prime)\propto{{\rho}\over{(M/L)_U}}\Biggl[\int_0^{z_l}
[\theta D(z)]^2{{cdt}\over{dz}}(1+z)^2 dz 
+\int_{z_l}^{z_{max}^\prime}
[\theta D(z)]^2{{cdt}\over{dz}}(1+z)^2 {1\over{A(z_l,z)}} dz\Biggr].
\eqno(4)$$
Average magnification due to the clump at $z_l$ of sources at $z>z_l$ is
$$A^{-1}(z_l,z)\approx1-2{{\Sigma_c}\over{\Sigma_{crit}}}=
1-{{2C(1+z_l)^2[\theta D(z)]^{-\beta}}\over{\Sigma_{crit}}}.\eqno(5)$$

Let us estimate $z_{max}^\prime$, the redshift of the faintest observable
GRBs if lensing magnification is included. The minimum observable 
(unlensed) flux is reduced to  
$f_{min}^\prime=f_{min} A^{-1}(z_l,z_{max})\approx f (1-2\kappa)$. 
The fractional change in flux is then $df/f\approx
-2\kappa$, and since ${{df}\over f}={{df}\over{dz}}{{1}\over f}dz$,
the redshift change is 
$$dz\approx 2\kappa {{(w^{0.5}-1)}\over{(a-2)w^{-0.5}-(a-1)w^{-1}}}
=x(w)\kappa,\eqno(6)$$
where $w=1+z_{max}$, and ($\Omega$, $\Lambda$)=(1, 0) cosmology is
assumed. With weak lensing, maximum redshift of observed bursts is 
$z_{max}^\prime=z_{max}+dz$.
For $z_{max}\sim 1$, and $a$ between 0.5 and 1.5, $x(w)$ is 
of the order of 1.

Now let us examine the effects of weak lensing due a single clump
at $z_l$, on the observed rate of GRBs seen in a given direction.
An interesting quantity to consider is the ratio of the 
change in the number of bursts due to lensing to the number of bursts 
due to the clump itself, $R(z_l)={{(F_l-F_u)}\over{F_c}}$. Making use of
equations (2)-(6) we get,
$$R(z_l)\approx\rho{{(M/L)_c}\over{(M/L)_U}}(1+z_l){1\over{D(z_l)^2}}
\Biggl[x(w){{D(z_{max})^2(1+z_{max})^2}\over{\Sigma_{crit}(z_l,z_{max})}}
{{cdt}\over{dz}}{\Bigg\vert}_{z_{max}}{\hskip 0.5in}$$
$${\hskip 3.0in} 
-{\int_{z_l}^{z_{max}}{{D(z)^2(1+z)^2}\over{\Sigma_{crit}(z_l,z)}}
{{cdt}\over{dz}}dz}\Biggr].\eqno(7)$$
The two terms in equation (7) represent additional sources due to 
fainter observable fluxes, and the deficit in the source number 
density due to the dilution effects of lensing, respectively. Ratio 
$R(z_l)$ can be either positive or negative depending on which of the 
two terms in equation (7) dominates. In general, area dilution term 
dominates if the column of matter between the lens and the edge of the 
source population is large. 

$R(z_l)$ is independent of $\beta$, $\theta$, and $C$, as long as lensing
is weak. In fact, it only depends on cosmology through $D$'s,
$\rho$, and $(M/L)$'s, and weakly, on $a$, the spectral index of GRBs.

Figure 1 shows $R(z_l)$ as a function of log~$z_l$, for ($\Omega$, 
$\Lambda$)=(1, 0).  There are five solid curves; from the top they are 
for $z_{max}=$0.5, 0.2, 0.7, 1.0, and 1.5. These curves
assume ${(M/L)_c}={(M/L)_U}$, \ie the density inhomogeneities on 
relevant scales, $>10h^{-1}$Mpc, have a biasing factor 
$b={{(M/L)_U}\over{(M/L)_c}}=1$. The long-dash and short-dash curves
are for $z_{max}=$1.0, and $b=2$ and $b=0.5$, respectively. Since 
$R(z_l)$ is simply proportional to ${{(M/L)_c}\over{(M/L)_U}}=b^{-1}$,
the rest of the $z_{max}$ cases with $b\ne1 $ are not shown on the plot 
to avoid crowding.
The actual value of $b$ is probably somewhere between 0.5 and 2, and 
depends on the type of galaxies that are used to delineate the density
inhomogeneities. For example, the biasing factor
for IRAS galaxies is probably around 0.7, while optical galaxies have a
larger $b$ (see Dekel \etal 1993, Strauss \etal 1992).

Notice that $R(z_l)$ changes very rapidly as redshift $z_l$ decreases 
(Figure 1 is a log-linear plot). In fact, the analytical expression in 
equation (7) diverges as $1\over {z_l}$, as $z_l$ approaches 0. 
Physically, the quantity $R(z_l)$ loses its meaning as $z_l$ gets very
small---since the expected number of bursts is well below 1. 

Figure 1 implies that weak lensing by the nearby cosmological structures
has a non-negligible effect ($\vert R\vert $ of the order of 1) on the 
distribution 
of GRBs of all fluxes. For example, if $z_{max}=1$, the excess of 
bright bursts due to the cosmological structures at $z_l\sim 0.015$ 
(log~$z_l\sim -1.82$) will be almost exactly offset by the deficit of 
fainter bursts. In other words, if GRBs of all fluxes are
considered, there will be no net change of the burst rate due to 
structures at $z_l\sim 0.015$.
Of course, with the present numbers of bursts in the BATSE
Catalog ($\sim 1000$), and the fraction of bursts originating at
$z_l\simlt 0.015$ ($\sim 1.6\cdot 10^{-5}$), one would not expect to see
either these structures, or their lensing effects.

To gauge the overall importance of weak lensing, one needs to consider
lenses at all redshifts up to $z_{max}$. Since $R(z_l)$ is independent of
the particulars of cosmological density inhomogeneities, a rough estimate 
of the average value of $R(z_l)$ is its volume weighted average, 
$$\langle R\rangle=\int_0^{z_{max}} R(z_l){{dV}\over{dz}}dz {\Bigg/}
\int_0^{z_{max}} {{dV}\over{dz}}dz,\eqno(8)$$
where $dV$ is the comoving volume element.

Figure 2 shows the dependence of $\langle R\rangle$, expressed as a
percentage, on the effective 
depth of the observed GRB population, $z_{max}$, for three different
sets of cosmological parameters: ($\Omega$, $\Lambda$)=(0.2, 0.8), 
(0.2, 0), and (1, 0). As expected, a $\Lambda$ dominated Universe
shows the largest change due to lensing.  

Volume weighted $\langle R\rangle$
is strongly dominated by the positive $R$'s at $z_l\simgt 0.1$, 
for all values of $z_{max}$ and all sets of ($\Omega$, $\Lambda$).
This means that weak lensing will enhance the contrast of individual 
density inhomogeneities by $\langle R\rangle$. Therefore, the observed 
angular correlation function of {\it all} bursts will be increased by 
$\langle R\rangle$ on {\it all} angular scales. Given that the present 
day limits on $\omega(\theta)$ of GRBs are rather weak, an increase of 
$\simlt 35\%$ would not be detected, especially if the observed GRBs 
extend to large redshifts, $z_{max}\simgt 0.5$.

\section{Conclusions and Discussion}

In this paper we examined the effect of weak lensing on the angular
distribution of GRBs. Since the nature of and distance to GRBs is 
unknown, a set of assumptions about the GRB population had to be adopted
(see Section 2.1). These assumptions are common to most models of GRBs
currently discussed in the literature. In particular, all the bursts are
considered to be at cosmological distances, and associated with luminous 
matter. Another assumption crucial to the current discussion is that 
GRBs are standard candles, or at least have a narrow intrinsic 
luminosity function. The main conclusions are as follows.

The dominant effect of weak lensing is to extend the faint flux cutoff 
of the observed bursts. This wins over the area dilution effects of
lensing, and hence leads to an enhancement of the density contrast
of cosmological structures, as seen in the distribution of bursts of
all fluxes. The amplitude of the enhancement depends on the depth of
the present GRB Catalogs, the cosmological model adopted, and the
biasing factor, $b$ on scales $>~10h^{-1}$ Mpc. The amplitude is 
insensitive to the mass profile of density inhomogeneities, 
$\beta$, their absolute
mass density normalization, $C$, and the angular scale of 
observations, $\theta$. It is slightly sensitive to the assumed 
spectral index of GRBs, $a$. 

The overall importance of weak lensing is summarized in Figure 2.
$\langle R\rangle$ can be roughly 
interpreted as the change in the amplitude of the angular correlation 
function of GRBs. 
For a canonical set of assumptions, $z_{max}=1$ and 
$(\Omega$, $\Lambda)$=(1, 0), $\omega(\theta)$ is expected to be 
increased by $\sim 8\%$ on all angular scales. Such a change will not
be detected given the present numbers of cataloged bursts.

Recently, it was suggested by Lamb and Quashnock (1993) that
if GRBs trace luminous matter, their distribution on $\simgt~1^\circ$
scales can be used to probe the large scale structure in the Universe. 
The authors estimate  
that if $\simgt 3000$ GRBs are observed, these would provide the 
power spectrum of density fluctuations on scales currently probed by
pencil beam surveys and superclusters. Density fluctuations on such
scales are probably primordial in nature and may be compared to those
implied by the COBE observations of microwave background.
If, after a few more years of BATSE observations, the accumulated number 
of bursts makes such an investigation possible, then 
weak lensing effects should be taken into account.

Since the weak lensing scenario described in this paper seems to be 
rather general, one may wonder if it applies to populations of other 
extragalactic objects seen in projection, for example, galaxies. 
The answer is no. Here, GRBs are treated as standard candles. Galaxies,
on the other hand, have a broad distribution of intrinsic luminosities.
In fact, a typical $d$log~$N(m)$/$dm$ slope for galaxies, 0.4 (Jarvis
\& Tyson 1981),
leads to no net change in the surface number density of a lensed 
population, hence the angular correlation function of galaxies will
not be affected by weak lensing in this case. 
(For an extended discussion of the effect of weak lensing
on the observed angular clustering of galaxies see Villumsen 1995.)

\acknowledgments

I would like to acknowledge the support of the PPARC fellowship 
at the Institute of Astronomy, Cambridge, UK. I am also grateful to the
anonymous referee for useful comments on the paper.

\clearpage

\figcaption{The effect of a single mass clump at $z_l$ on the number 
density of bursts seen in that direction. $R$ is ratio of the
change in the number of bursts due to lensing by a clump at $z_l$, 
to the number of bursts due to the clump itself.
See equation (7). Solid curves are for an $\Omega$=1 Universe and the 
biasing factor of galaxies,  $b$=1. Solid curves are labeled by the 
maximum redshift of observable bursts, $z_{max}$. Long-dash and 
short-dash curves are for $z_{max}$, but with $b$=2 and 0.5, 
respectively.}

\figcaption{The collective effect of all density inhomogeneities on the
angular distribution of the bursts. $\langle R \rangle$ is the average 
volume weighted value of $R(z_l)$, expressed as a percentage. See 
equation (8).  $\langle R \rangle$ can be roughly interpreted as the 
fractional enhancement in the angular correlation function of all GRBs 
on all angular scales, due to weak gravitational lensing
Three sets of cosmological parameters are considered, as indicated 
by the labels on the plot.}

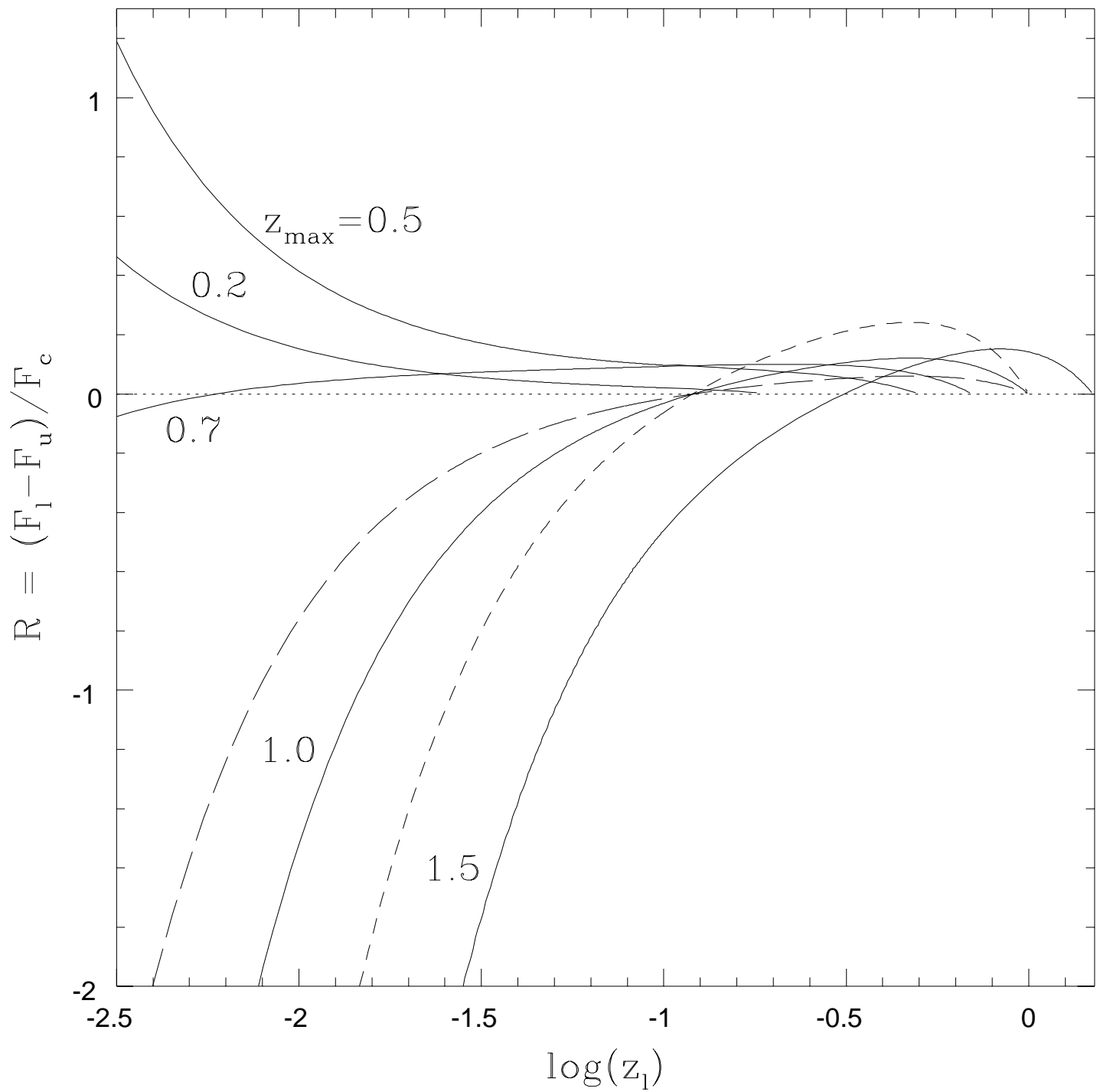
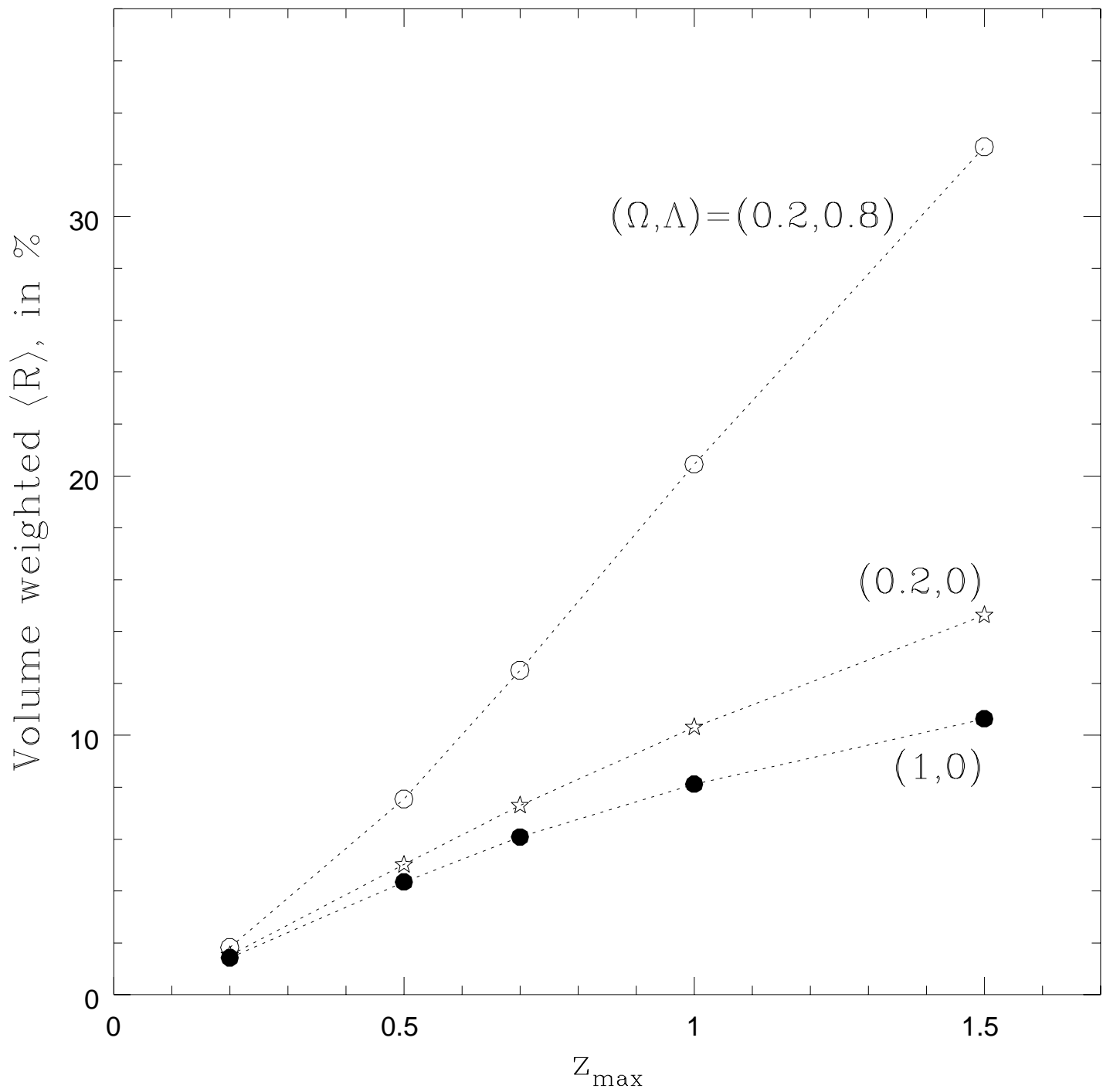
\end{document}